\documentclass[sn-basic]{sn-jnl}

\usepackage{makecell}	
\usepackage{natbib}	
\usepackage{tabularx}	

\theoremstyle{thmstyleone}%
\theoremstyle{thmstyletwo}%

\theoremstyle{thmstylethree}%

\makeindex
\begin{document}

\title[]{Detecting and analyzing missing citations to published scientific entities}


\author[1]{\fnm{Jialiang} \sur{Lin}}\email{me@linjialiang.net}

\author[1]{\fnm{Yao} \sur{Yu}}\email{eloria1995@gmail.com}

\author[1]{\fnm{Jiaxin} \sur{Song}}\email{songjiaxin@stu.xmu.edu.cn}

\author*[1]{\fnm{Xiaodong} \sur{Shi}}\email{mandel@xmu.edu.cn}

\affil[1]{\orgdiv{School of Informatics}, \orgname{Xiamen University}, \orgaddress{\city{Xiamen}, \country{China}}}


\abstract{Proper citation is of great importance in academic writing for it enables knowledge accumulation and maintains academic integrity. However, citing properly is not an easy task. For published scientific entities, the ever-growing academic publications and over-familiarity of terms easily lead to missing citations. To deal with this situation, we design a special method Citation Recommendation for Published Scientific Entity (CRPSE) based on the cooccurrences between published scientific entities and in-text citations in the same sentences from previous researchers. Experimental outcomes show the effectiveness of our method in recommending the source papers for published scientific entities. We further conduct a statistical analysis on missing citations among papers published in prestigious computer science conferences in 2020. In the 12,278 papers collected, 475 published scientific entities of computer science and mathematics are found to have missing citations. Many entities mentioned without citations are found to be well-accepted research results. On a median basis, the papers proposing these published scientific entities with missing citations were published 8 years ago, which can be considered the time frame for a published scientific entity to develop into a well-accepted concept. For published scientific entities, we appeal for accurate and full citation of their source papers as required by academic standards.
}




\maketitle

\section{Introduction}
\label{sec:introduction}

Academic research is always in development and never in isolation, as it has to build upon the existing knowledge to create new knowledge, which then becomes the foundation for future research. Citations in the end product of research, i.e.\ the research papers, are the embodiment of this connection between the current research and its predecessors. Without this connection, there will be no creation and accumulation of knowledge. Moreover, citations also put one's work into an academic context. On one hand, it bolsters readers' understanding of the current work by preparing readers with adequate information on a certain field. Further, it strikes a common ground between the author and the reader. On the other, citations can verify the credibility of the current work. Proper citations show that the author is well-informed of this particular field. Others' work can also be evidence to support the author's arguments. Therefore, the importance of appropriate citing can not be overestimated, but citing accurately and fully is a difficult task~\citep{hicks-how-2021}.

In the first place, with academic publications in overwhelming abundance, it is hard to decide which one is the right source to cite. The publishing industry has been digitized in general and academic publishing is no exception. It has become much more easier for knowledge to propagate when stored at a digital platform. As stated by the Dimensions database,\footnote{https://app.dimensions.ai/discover/publication.} the last decade, from 2011 to 2020, witnesses a continuous growing trend in academic publishing, with a 74.9\% growth rate.\footnote{Data was obtained in November of 2021.} Apart from official publications, digital repositories for e-prints, like arXiv~\citep{ginsparg-winners-1997}, also register an upward movement in submissions, with 77.1\% submissions of computer science on arXiv becoming peer-reviewed publications eventually~\citep{lin-how-2020}. The rapid increase in the amount of published information creates enormous difficulties for researchers to manage data available to them, and thus causes the problem of information overload~\citep{gross-the-1964,roetzel-information-2019}. For any researcher, time and energy are limited resources that need to be managed efficiently. At the same time, keeping up to date with newly published research in a certain field is crucial as it prevents one's work from being repetitive or derivative. With the sheer number of online services and platforms available to researchers, it becomes unprecedentedly difficult to do a thorough literature search among the overloaded academic data. Worse still, academic information overload also causes inappropriate citation practices. Researchers can make mistakes in citing as they do not have enough time to learn more about the citing source, or they might just rely on secondary sources without reading the source papers.

With new knowledge come new published scientific entities, such as PageRank~\citep{page-the-1999}, MapReduce~\citep{dean-mapreduce-2004} and Transformer~\citep{vaswani-attention-2017}, which serve as symbols for research. These entities, newly coined by the researchers to represent their ideas and findings, are generally introduced in papers. Citing these entities consistently and accurately is important in academic writing, as it makes academic connections and maintains academic integrity. However, it is not uncommon for inappropriate citations of these entities to occur. Among the published scientific entities with inappropriate citation practices, software and frameworks are highly inclined to be used in academic papers without proper citations. Some software and frameworks are introduced officially in published papers, but many researchers only conveniently cite their web pages. TensorFlow is a typical example. With its first version released in 2015, TensorFlow has become a prestigious and widely used framework for deep learning. Some researchers only cite its official website\footnote{https://www.tensorflow.org/.} or its GitHub page\footnote{https://github.com/tensorflow.} when using TensorFlow in their papers. In fact, the TensorFlow team has published a paper ``TensorFlow: A System for Large-Scale Machine Learning''~\citep{abadi-tensorFlow-2016} to introduce this framework in 2016. With the publication of this paper, the proper citation of TensorFlow in academic papers shall be this paper rather than the web pages. Yet this paper has been ignored by some researchers, in part because they fail to catch up with its publication.

Furthermore, some researchers are unaware of missing citations. Citations build up the edifice of knowledge by turning factual claims into what eventually being accepted as established knowledge over time. The existence of some published scientific entities has been regarded as usual by some researchers, after being cited over and over again in the publications. For these entities, it is more and more common that no citations are provided, which does not follow the academic norms.

We can use BLEU~\citep{papineni-bleu-2002} and SciBERT~\citep{beltagy-scibert-2019} in comparison as a distinct case in point. BLEU is a method for automatic evaluation of machine translation proposed in 2002. This method soon becomes and still is the most widely used metric. However, the more well-known it is, the less proper citations it has. In our research, we find that a large amount of recent published papers mentioning this method fail to attribute it to its source. It seems that the missing citation does not hinder readers' understanding, for BLEU is basic knowledge that researchers in a related domain should be familiar with. But citing the source paper of BLEU helps to increase the credibility of papers and it is a behavior to credit and honor the authors who proposed it at the very beginning. It also helps new researchers to develop a good grasp of the root of this mature concept. In comparison, the newly introduced entity SciBERT stands in striking contrast to BLEU. Proposed in 2019, SciBERT misses no citation among the papers selected for the experiments in Sect.~\ref{sec:detect}. As can be seen from here, time-honored research with important results is more likely to be used without proper citations.

In summary, citing properly of published scientific entities is no easy task and can create an extra burden to academic writing. To address this issue, we propose a method to map published scientific entities to their source papers. Our main contributions lie in:

\begin{itemize}

\item We provide a detailed approach to construct a large-scale published scientific entity-papers mapping dataset;

\item We propose a local citation recommendation method Citation Recommendation for Published Scientific Entity (CRPSE) based on the dataset above. The method is proved to produce great performance in the task of citation recommendation for published scientific entities;

\item We conduct an extensive statistical analysis on published scientific entities with missing citations among papers published in prestigious computer science conferences in 2020 with the employment of the method above.

\end{itemize}

\section{Related work}
\label{sec:relatedwork}

Our research is closely related to the following two fields.

\subsection{Local citation recommendation}

First raised by \citet{strohman-recommending-2007}, citation recommendation is ``the problem of academic literature search by considering an unpublished manuscript as a query to a search system''. It is later divided into global citation recommendation and local citation recommendation by \citet{he-context-2010}. \citet{he-citation-2011} first introduce a method to localize in-text citations automatically and \citet{huang-recommending-2012} later incorporate machine translation with this method for improvement. \citet{huang-neural-2015} propose a neural probability model for learning feature representation of distributed word embeddings, and predict the citation probability for papers on the grounds of semantic distance through a multi-layer neural network. \citet{chen-citation-2019} describe a citation recommendation algorithm CIRec based on citation tendency, which incorporates a weighted heterogeneous network, a biased random walk procedure and the skip-gram model~\citep{mikolov-efficient-2013}. In recent years, with the development of deep learning, complex neural networks are used in local citation recommendation. \citet{ebesu-neural-2017} present Neural Citation Network (NCN) constructed on the encoder-decoder architecture. Convolutional neural networks (CNN)~\citep{lecun-backpropagation-1989,yin-personalized-2017}, long short-term memory (LSTM)~\citep{hochreiter-long-1997,yang-lstm-2018} and graph convolutional networks (GCN)~\citep{kipf-semi-2017,jeong-context-2020} are more examples of successful deep learning-based local citation recommendation. In these methods, deep learning is used to learn the text feature, which greatly enhances the understanding and feature representation of text.

\subsection{Citation data analysis}

Many research efforts have been devoted to citation data analysis. Some focus on analyzing the characteristics of the papers cited and the influence of different citation behaviors on the cited papers. \citet{hu-understanding-2017} study the repetitive mentioning of a reference in the citing paper. Self-citation is a popular topic in citation data analysis. It is found to have a positive impact on the cited paper~\citep{fowler-self-2007}. This is also proved in \citet{amjad-scientific-2020}'s work. They find in the analysis of five self-citation trends that papers without self-citation tend to have dampened research impact. \citet{yan-authors-2020} discover a moderate increase in citations of papers after the authors of which are awarded Nobel Prize. Attention is also paid to the research of dataset citation. \citet{zhao-data-2018} conduct a content analysis to study the use of datasets in different disciplines. They find that researchers tend to build their own datasets instead of reusing the existing datasets. Some studies of missing citations target for patents and articles. \citet{oh-recommending-2014} propose a missing citation recommendation method for new patents utilizing the patent citation network. \citet{ciotti-homophily-2016} design a bibliography-based recommendation method, which quantifies the similarity between articles and further to uncover missing references. This bibliography-based recommendation method is also found to facilitate the dissemination of scientific results.

\section{Method}
\label{sec:method}

In this section, we describe our method of Citation Recommendation for Published Scientific Entity (CRPSE) in detail. We first give the definition of terms and the problem. Then we provide the step-by-step construction process of the published scientific entity-papers mapping dataset which is used as the basis for our citation recommendation method. The process mainly contains two steps: collecting and filtering. After that, we present two sorting criteria to put the candidate source papers in order with the support of the constructed dataset. Last but not least, we delineate the recommendation pipeline with an example.

\subsection{Definition of terms}

\begin{itemize}

\item \textbf{Paper}

A paper is defined as an academic publication in general. It is not restricted to journal or conference papers, but it can also refer to books, reports, patents and standards.

\item \textbf{Published scientific entity}

A published scientific entity is defined as a scientific object that has a published paper deriving from it and the author of the paper is also the researcher that proposes this scientific object. The published scientific entities can be categorized into two types depending on how they are named. The first type is named by the original author, such as the neighbor-joining method~\citep{saitou-neighbor-1987}, ATRP~\citep{wang-controlled-1995}, ImageNet~\citep{deng-imagenet-2009} and AlphaGo~\citep{silver-mastering-2016}. The second type is named later by other researchers, usually named after the original author. Examples include Schrödinger equation~\citep{schrodinger-undulatory-1926}, Bradford's law~\citep{bradford-sources-1934}, Turing test~\citep{turing-computing-1950} and Witten-Bell smoothing~\citep{witten-zero-1991}. Published scientific entity is also referred to as published entity in this paper.

\end{itemize}

\subsection{Problem definition}

$p$ is a paper requiring citation recommendation. $s_i$ denotes a sentence in $p$, $s_i \in S=\{s_1,s_2,\ldots,s_I\}$, where $I$ is the total number of sentences in $p$. $e_j$ denotes a published entity in $p$, $e_j \in E=\{e_1,e_2,\ldots,e_J\} (J > 0)$, where $J$ is the total number of published entities in $p$. When $J = 0$, $E$ is $\varnothing$. $d_j$ denotes the source paper that proposes $e_j$. The problem of CRPSE is defined as processing the sentence set $S$ in $p$ to build the entity set $E$ and finding an ordered set $C_j$ consisted of $K$ relevant papers for each entity $e_j$.

\begin{align*}
& C_j=\{c_{j_1},c_{j_2},\ldots,c_{j_K}\}  \\
& \text{s.t.} \quad r_{j_1} > r_{j_2} > \ldots > r_{j_K}  \\
\end{align*}

\noindent
$K$ is a parameter that can be set as required. $c_{j_k}$ is one candidate source paper of $e_j$ and $r_{j_k}$ denotes the score of $c_{j_k}$ used in ordering, $k \in \{1,2,\ldots,K\}$. The objective of CRPSE is to let $d_j$ belong to $C_j$ and its corresponding ordering score be the highest. That is:

\begin{align*}
d_j = c_{j_1} \in C_j
\end{align*}

\subsection{Constructing the published scientific entity-papers mapping dataset}

To achieve the objective of CRPSE, we track the cooccurrences between published entities and in-text citations in the citing behaviors of researchers. Based on the cooccurrences, we analyze a large quantity of citations and construct a published scientific entity-papers mapping dataset as the base to guide our own citation recommendation. Semantic Scholar Open Research Corpus (S2ORC)~\citep{lo-s2orc-2020} is used to construct this dataset. S2ORC is by far the largest public corpus of English academic papers. The corpus covers 81.1 million papers from various domains with rich metadata, including titles, abstracts, authors, publication year, references and full text of 8.1 million open access papers. In S2ORC, in-text citations of these open access papers are annotated and linked to their resolved references, which creates 380.5 million pieces of citation data. We take great advantage of this corpus, mining citation information from its data to construct our dataset. It is important to note that the construction method proposed here is not specially designed for S2ORC. It is a universal method and can be transferred to any other corpus with in-text citation markers.

\begin{figure}[ht]
\centerline{\includegraphics[width=0.9\textwidth]{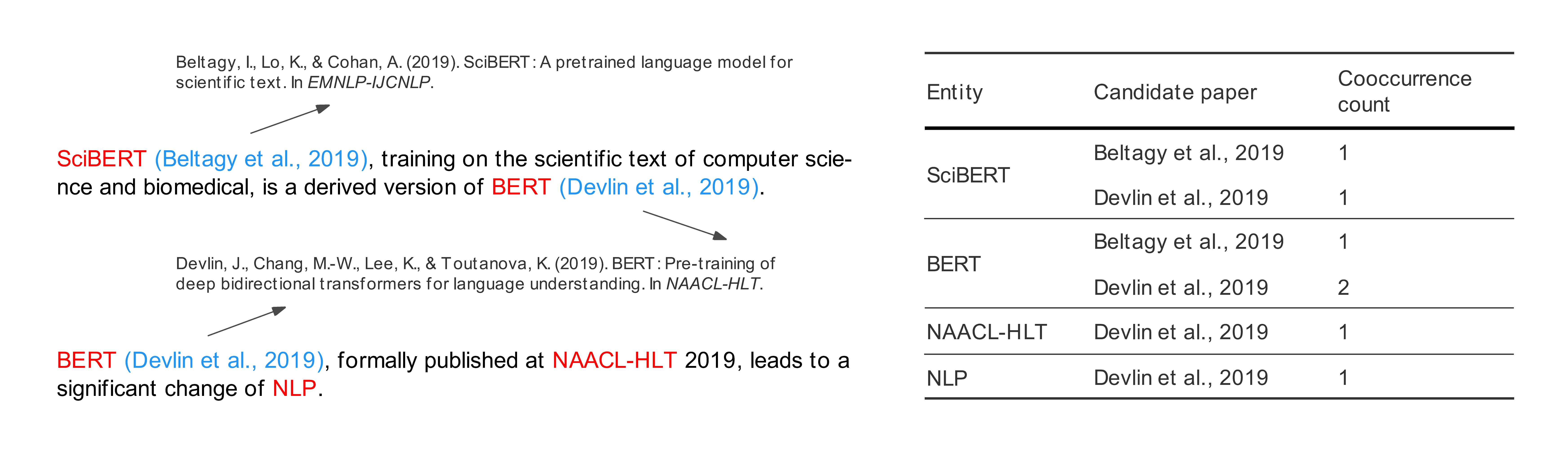}}
\caption{Illustration of the mapping between entities and in-text citations}
\label{fig:illustration-collecting}
\end{figure}

The construction of our published scientific entity-papers mapping dataset is elucidated as follows and is illustrated in Fig.~\ref{fig:illustration-collecting}.\footnote{The examples in the figure are created for better illustration, not real examples from S2ORC.} First, we take each paper with both full text and annotated resolved bibliographic references provided by S2ORC and segment its main text into sentences using scispaCy~\citep{neumann-scispacy-2019}. ScispaCy is a powerful Python library developed from spaCy\footnote{https://github.com/explosion/spaCy.} for scientific text processing. It is at an industrial-strength level for tagging, parsing, entity extraction and other tasks in natural language processing. We use scispaCy for sentence segmentation and entity extraction in this paper. Second, for each general entity $e_{gen_j}$ (words in red color) detected by scispaCy in every segmented sentence, we take all papers with their in-text citations (words in blue color) cooccurring in the same sentence with $e_{gen_j}$ as the candidate source papers for $e_{gen_j}$. One candidate source paper is denoted by $c_{j_l}$, $c_{j_l} \in C_j=\{c_{j_1},c_{j_2},\ldots,c_{j_L}\} (L > 0)$, where $L$ is the total number of candidate source papers of $e_{gen_j}$. When $L = 0$, $C_j$ is $\varnothing$. The cooccurrence count of $e_{gen_j}$ and $c_{j_l}$ is denoted by $n_{j_l}$. With each cooccurrence, the value of $n_{j_l}$ will be added 1.

In the left part of Fig.~\ref{fig:illustration-collecting}, there are two sample sentences with five entities, three in-text citations and two corresponding papers. After being processed following the steps described above, a dataset can be generated as shown in the right part of Fig.~\ref{fig:illustration-collecting}.

All papers with both full text and annotated resolved bibliographic references in S2ORC are processed following the steps above to construct a raw published scientific entity-papers mapping dataset. Some entities and their candidate source papers only share a small number of cooccurrences, which implies that they do not have a strong correlation. So an entity $e_{gen_j}$ without a candidate source paper with a cooccurrence count equal to or greater than 20 will be removed from the dataset. That is $n_{j_l} < 20, \forall l \in \{1,2,\ldots,L\}$.

This raw dataset is full of outliers that need to be removed. ScispaCy detects all entities, including some non-published entities, like NAACL-HLT and NLP in Fig.~\ref{fig:illustration-collecting}. The non-published entities are outliers should be cleared out. We train a binary classification model to separate the outliers from published entities in the raw dataset.

The positive and negative samples for our classification model are created as follows with both positive and negative samples belonging to the published scientific entity-papers mapping dataset. The positive samples are outliers that should be filtered out. They are composed of high-frequency surnames in S2ORC and high-frequency words. We compile a list of all author surnames in S2ORC and select the top 10,000 high-frequency surnames as positive samples. For high-frequency words, we pick out the top 14,000 words from English Word Frequency dataset\footnote{https://www.kaggle.com/rtatman/english-word-frequency.} with high-frequency surnames screened out. In total we have 24,000 positive samples. The negative samples are published entities that should be retained in the dataset. To locate such entities, we study a large quantity of academic papers' titles. Through a close study, we find a structural pattern shared by these titles. In most titles, if there is a colon used in the title with only one single uncommon word before the colon, then this word is likely to be a published entity that we are looking for to construct the negative set. Example titles include ``AllenNLP: A Deep Semantic Natural Language Processing Platform''~\citep{gardner-allennlp-2018} and ``VoxCeleb2: Deep Speaker Recognition'' ~\citep{chung-voxceleb2-2018}. In these two examples, the words ``AllenNLP'' and ``VoxCeleb2'' before the colon are the published entities for the negative set. We process all the titles of S2ORC collected papers that follow this pattern and gather the word before the colon to construct the negative set. After those words were collected, we perform filtering to ensure they are not the author surname in S2ORC and are not in the English Word Frequency dataset. Additionally, we manually collect other published entity samples from Computing Research Repository (CoRR)~\citep{halpern-corr-2000} manuscripts on arXiv submitted between May 2020 and July 2020. In the end, we have 12,000 negative samples.

\begin{figure*}
\centerline{\includegraphics[width=0.9\textwidth]{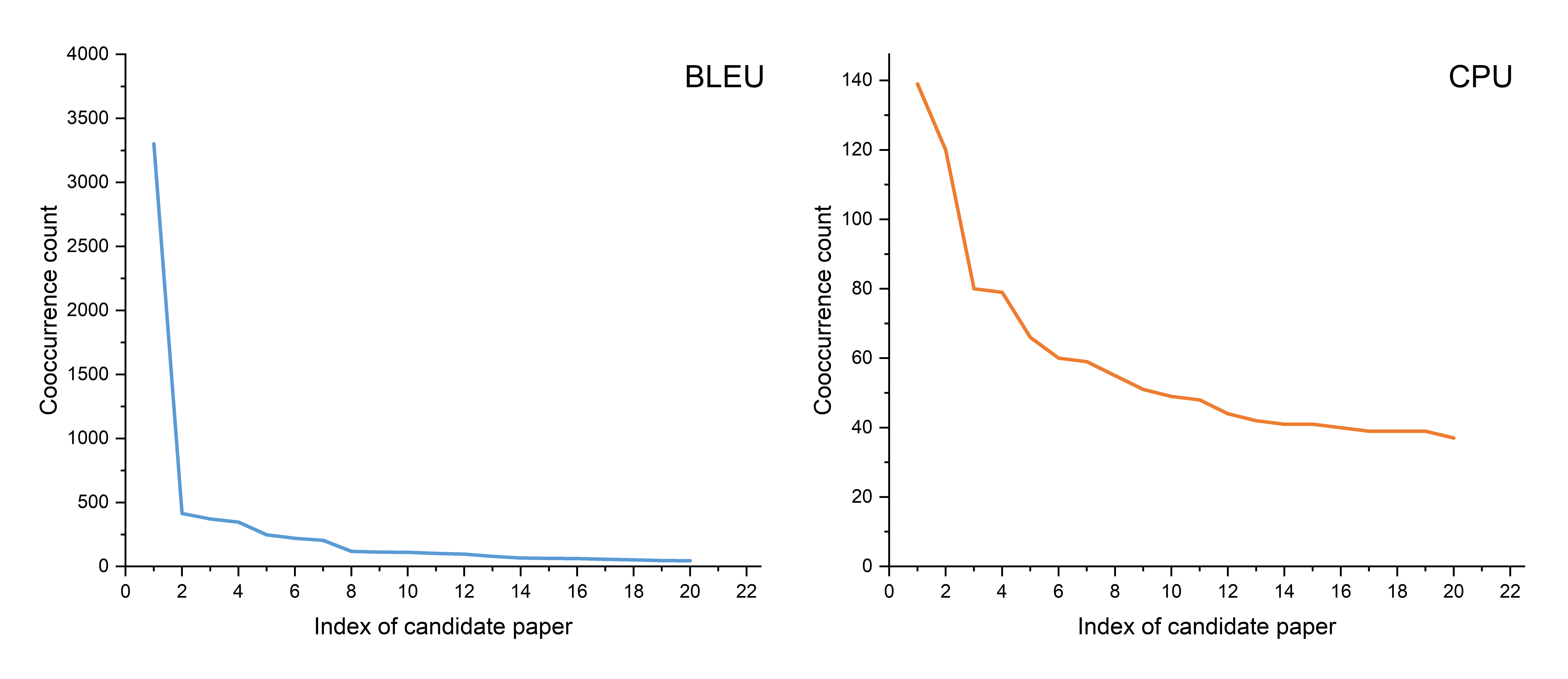}}
\caption{Comparison of the cooccurrence distribution between a published scientific entity (BLEU) and an outlier (CPU)}
\label{fig:normal-outlier-entity}
\end{figure*}

After constructing the two sample sets, we progress to selecting the feature for classification. We sort and convert all the cooccurrence counts of candidate source papers of an entity into a line graph. By looking at the line graphs, we find two different patterns in the distribution of the cooccurrence count. For outliers, the curves in the line graph decrease very slowly. For published entities, the curves present dramatic falls. See Fig.~\ref{fig:normal-outlier-entity}.

It is not hard to interpret the difference between the cooccurrence count of outliers and that of published entities. For outliers, the candidate source papers share a similar cooccurrence count, because these outliers, like high-frequency surnames and high-frequency words, have no specific papers that propose them. In contrast, for published entities, there will be one candidate source paper that has a significantly higher cooccurrence count. This is the source paper that introduces the published entity and most researchers cite this paper when mentioning the entity, hence the distinct contrast in cooccurrences between the source paper and the rest. On the basis of the analysis above, the cooccurrence counts are used as the classification feature.

We divide the overall sample set randomly into a training set, a validation set and a test set at the ratio of 8:1:1. We test the performance of different classifiers and random forests~\citep{breiman-random-2001} is chosen accordingly for this task. It is an ensemble classifier that combines a number of decision trees and the result of final classification is determined by voting of each sub-tree. This classifier has good performance while controlling over-fitting. The F1-score of the trained model on the test set is 0.833. We use the trained model to filter out the outliers. After removing the outliers, we have hundreds of thousands pieces of published entities with candidate source papers and their cooccurrence counts in the published scientific entity-papers mapping dataset.

\subsection{Sorting criterion}

For each published entity, there are numerous candidate source papers. Those papers need to be sorted by certain criteria to produce the final result of recommendation. For this purpose, two sorting criteria are designed.

\begin{itemize}
\item \textbf{Cooccurrence count-based sorting criterion}

This is a simple and straightforward method. Each candidate source paper of an entity is given a cooccurrence count (see Fig.~\ref{fig:illustration-collecting}). This cooccurrence count indicates how many times a paper is presented as an in-text citation in the same sentence with this entity in S2ORC. It is a strong quantitative index of the correlation between an entity and a paper. The higher the cooccurrence count, the stronger the correlation between the entity and the candidate source paper, i.e.\ higher possibility for this paper to be the entity's source paper. On the basis of this positive correlation, a cooccurrence count-based score is computed to be the first sorting criterion. This scoring makes full use of the citation behavior of researchers, looking into what is most-cited when the researchers mention this entity. The formulation of the cooccurrence count score of the $l$-th paper $r_{count_{l}}$ in the mapped candidate source paper set of a published entity $e$ is as follows:

\begin{equation}
r_{count_{l}} = w_{count_{l}} = \frac{n_l}{SC}
\end{equation}

\begin{equation}
SC = \sum_{l=1}^{L} n_l
\end{equation}

where $n_l$ is the cooccurrence count of a candidate source paper $c_l$ for $e$ and $L$ is the total number of candidate source papers of $e$. $SC$ is the sum of cooccurrence counts that $e$ has with all the candidate source papers.

\item \textbf{Weighted context embedding-based sorting criterion}

The first method makes use of the positive correlation of cooccurrence between an entity and its candidate source papers, but it does not take into consideration what the entity is and in which context it is mentioned. It can sort out the candidate source papers in order correctly for the majority of published entities, but not for polysemous entities, i.e.\ entities that have a name with more than one meaning. SAFD is one example. It can mean Stirring As Foam Disruption~\citep{hoeks-stirring-1997}, Statistical Adaptive Fourier Decomposition~\citep{tan-new-2019} or Single Shot Anchor Free Face Detector~\citep{wang-safd-2021}. It relies largely on the context in which the entity is used to decide its contextual meaning and recommend its source paper. To address the problem of polysemy, we use context embedding to find out the contextual correlation between an entity and its candidate source papers. The computation is shown as follows:

\begin{equation}
w_{context_{l}} = CosSim(SciBERT(s),SciBERT(t_l))
\end{equation}

where $SciBERT$ denotes the method of generating embedding using SciBERT~\citep{beltagy-scibert-2019}. $s$ is the sentence in which the published entity $e$ is located. $t_l$ is the text that concatenates the title and abstract of the $l$-th candidate source paper of $e$ in its candidate source paper set. $CosSim$ denotes the calculation of cosine similarity. Then the weighted context embedding score of the $l$-th candidate source paper $r_{mix_{l}}$ can be calculated as the formulation below:

\begin{equation}
r_{mix_{l}} = \lambda w_{count_{l}}+(1-\lambda)w_{context_{l}}
\end{equation}

To improve the performance of this criterion, the hyperparameter $\lambda$ is applied.

\end{itemize}

\subsection{Recommendation pipeline}

The pipeline of CRPSE is as follows. We use scispaCy to detect entities in a paper. For each entity detected, we determine whether this entity is a published entity based off checking whether it is collected in the published scientific entity-papers mapping dataset. If yes, we sort its candidate source papers in order by one of the two sorting criteria above and select the top $K$ papers as the final recommended source papers. $K$ is usually set as 1, 5 or 10.

We use the sentence in Fig.~\ref{fig:illustration-collecting}, ``BERT, formally published at NAACL-HLT 2019, leads to a significant change of NLP'' as an example of using CRPSE for citation recommendation. First, using scispaCy, entities BERT, NAACL-HLT and NLP are detected. Through the checking in the constructed published scientific entity-papers mapping dataset, only BERT is included in this dataset. Therefore, only the entity BERT will be passed to the next step for recommendation. When $K$ is set to be 5 which means 5 papers are required to be recommended and the cooccurrence count-based sorting criterion is used, the recommended papers for the entity BERT in order are ``BERT: Pre-Training of Deep Bidirectional Transformers for Language Understanding''~\citep{devlin-bert-2019}, ``Deep Contextualized Word Representations''~\citep{peters-deep-2018}, ``Attention Is All You Need''~\citep{vaswani-attention-2017}, ``Improving Language Understanding by Generative Pre-Training''~\citep{radford-improving-2018} and ``XLNet: Generalized Autoregressive Pretraining for Language Understanding''~\citep{yang-xlnet-2019}. Among these papers, the top one is the source paper proposing the entity BERT and the other four papers are also related to the entity BERT.

\section{Experiment}
\label{sec:experiment}

In this section, we conduct an evaluation on the performance of CRPSE.

\subsection{Dataset}
\label{subsec:experiment-dataset}

To the best of our knowledge, there is no dataset specialized in evaluating the performance of local citation recommendation for published entities. So we manually construct a dataset for this task. We first download 101 manuscripts covering all categories under CoRR on arXiv submitted in October 2020. Then we screen out all general entities and the sentences they are in from the main text of those manuscripts. Published entities are labeled and their source papers are collected to build the final evaluation dataset.

\subsection{Result and analysis}

We use recall@1, recall@5, recall@10, mean average precision (MAP)~\citep{voorhees-overview-1998} and mean reciprocal rank (MRR)~\citep{voorhees-trec-1999} as metrics for the evaluation of CRPSE. Results are shown in Table~\ref{tab:method-performance}. As we can see from Table~\ref{tab:method-performance}, our method produces good performance as evaluated by all the metrics. The sorting criterion of weighted context embedding outperforms the criterion of cooccurrence count under the evaluation of most metrics. We have tried to compare our method with the state-of-the-art (SOTA) local citation recommendation methods, but the comparison is hard to conduct because of the low comparability between our method and other local citation recommendation methods. This low comparability is explained as follows:

\begin{itemize}

\item The comparison will be biased if we test our method and other general-purpose citation recommendation methods on the dataset constructed in Sect.~\ref{subsec:experiment-dataset}. This is because our method is designed for recommending citations for published entities, but as far as our literature search goes, none of the existing citation recommendation methods are specially aimed at published entity citation recommendation, hence the biased comparison.

\item The existing datasets for citation recommendation evaluation are not compatible with our citation recommendation method. These datasets are not designed for published entity citation recommendation and contain non-entity citation data other than entity citations. The performance of our method will be inappropriately compromised when tested on non-entity citation recommendation, for it is not the field for our method. In addition, some published entities are used without citations in these datasets. If using our method to recommend citations for such entities with poor citation practices, the recommendation will be mistakenly determined as wrong.

\item The published scientific entity-papers mapping dataset is not suitable for the training of some SOTA methods. The dataset is particularly designed for recommending citations for published entities. It does not contain paper metadata (e.g., published years, published venues and author information) for the reason that the use of CRPSE does not need such metadata. But this metadata is indispensable to some SOTA methods.

\end{itemize}

\begin{table}[ht]
	\caption{Evaluation results of Citation Recommendation for Published Scientific Entity (CRPSE)}
	\label{tab:method-performance}
	\centering
	\begin{tabular}{lll}
		\hline\noalign{\smallskip}
		\textbf{Metric}  &  \textbf{Cooccurrence count-based}  &  \textbf{Weighted context embedding-based} \\
		\hline\noalign{\smallskip}
		\textbf{Recall@1}   &  0.437           &  \textbf{0.499}   \\
		\textbf{Recall@5}   &  \textbf{0.783}  &  0.773   \\
		\textbf{Recall@10}  &  \textbf{0.804}  &  \textbf{0.804}    \\
		\textbf{MRR}        &  0.661           &  \textbf{0.698}     \\
		\textbf{MAP}        &  0.619           &  \textbf{0.649}    \\
		\hline\noalign{\smallskip}
	\end{tabular}
\end{table}

\subsection{Analysis of error types}

There are two typical error types found in our method.

\begin{itemize}

\item \textbf{Erroneous extraction}

We use scispaCy for entity extraction because it is a well-developed tool at an industrial-strength level. Nonetheless, extraction errors can still happen, especially in the processing of sentences with special symbols like mathematical expressions. We are looking for improvement in scispaCy to reduce these extraction errors.

\item \textbf{Over-screening}

The threshold for an entity in S2ORC to be included in our dataset is to have a cooccurrence count of 20 at least with any of its candidate source papers. Some published entities that are newly-proposed or less-mentioned are thus screened out by this threshold as these entities have rather low citations. We are working on a solution to address this problem of over-screening.

\end{itemize}

\section{Missing citation of published scientific entity}
\label{sec:detect}

The experiments above prove the great performance of CRPSE in mapping published entities with their source papers. We further conduct a statistical analysis on missing citations of published entities among computer science conference papers with this method applied to detect published entities without proper citations. Open access papers published in 14 prestigious computer science conferences in 2020 are collected for the analysis. Even though it is generally considered that conference papers tend to be less rigorous than journal papers, we still use these papers for analysis because of the following reasons: 1) conferences, especially top conferences in computer science attract more attention than journals~\citep{vrettas-conferences-2015}; 2) the conference papers we selected are published under open access license with more friendly accessibility.

These conferences (sorted by alphabetical order of their abbreviations) are the AAAI Conference on Artificial Intelligence (AAAI), the Annual Meeting of the Association for Computational Linguistics (ACL), the International Conference on Computational Linguistics (COLING), the Annual Conference on Learning Theory (COLT), the IEEE Conference on Computer Vision and Pattern Recognition (CVPR), the European Conference on Computer Vision (ECCV), the Conference on Empirical Methods in Natural Language Processing (EMNLP), the International Conference on Learning Representations (ICLR), the International Conference on Machine Learning (ICML), the International Joint Conference on Artificial Intelligence (IJCAI), the Conference of the International Speech Communication Association (INTERSPEECH), the Annual Conference on Neural Information Processing Systems (NeurIPS), the Robotics: Science and Systems (RSS) conference and the USENIX Security Symposium (USS).

\subsection{Definition}

Missing citation of published scientific entity is defined as inappropriate citing practice of mentioning a published entity in a paper without giving its source paper in the references. Citing a website or other source instead of the source paper that proposes this entity is still deemed as missing citation.

\subsection{DMC formulation}

For a given paper, the collection of all its reference papers is $REF$. A published entity in the given paper is denoted by $e_{j}$. The published scientific entity-papers mapping dataset constructed in Sect.~\ref{sec:method} is denoted by $DS$. $rec_{j_1}$ denotes the top 1 candidate source paper of $e_{j}$ with the highest weighted context embedding score recommended through CRPSE. The function of detecting missing citations (DMC) of $e_j$ is defined as follows:

\begin{equation*}
DMC(e_j)=\left\{
\begin{array}{rcl}
	& \{rec_{j_1}\} & \qquad {e_j \in DS \enspace and \enspace rec_{j_1} \notin REF}  \\
	& \varnothing & \qquad {e_j \notin DS}  \\
	& \varnothing & \qquad {e_j \in DS \enspace and \enspace rec_{j_1} \in REF}  \\
\end{array} \right.
\end{equation*}

\subsection{DMC pipeline}

First, we download all PDF files of regular papers published in the conferences mentioned above and convert them into XML files using GROBID~\citep{lopez-grobid-2009}. Second, we use scispaCy to segment their main text and detect entities. Third, for those entities, we get the results of CRPSE. Fourth, we extract references from the XML files. Fifth, the recommended papers are checked whether they are included in the references. If the recommended paper of a published entity is not in the references, the entity is considered as a potential entity with missing citation. These potential entities are double checked and confirmed if their recommended papers are not given in the references provided by Semantic Scholar API\footnote{https://api.semanticscholar.org/.} to prevent the possible parsing error of GROBID.

\subsection{Analysis}

The authenticity of the recommended papers needs to be confirmed to ensure the statistical analysis is accurate and valid. Taking into consideration our academic backgrounds, we only check and conduct the analysis on recommended papers from computer science and one of its most related domains, mathematics in our research.

Due to various writing requirements from different researchers, the same entity might have different forms of expression. For example, Adam algorithm, Adam method, Adam optimiser, Adam optimizer and Adam update rule all refer to the same published entity Adam~\citep{kingma-adam-2015}. To have a more accurate analysis result, different expressions of the same entity need to be merged into one. The entities with the same recommended paper are merged into a unified one by their longest common subsequences. Few cases are merged manually as they are hard to be merged with this method.

From these 14 computer science conferences, 12,278\footnote{Papers with parsing errors are excluded.} conference papers are collected. In these papers, 475 published entities in computer science and mathematics are found to have missing citations. This figure reveals that missing citations is not uncommon even in papers published in top computer science conferences.

In our analysis of the data, inaccurate citations of some published entities are found. For example, for the published entity DeepLabv3+ in one paper, ``Rethinking Atrous Convolution for Semantic Image Segmentation''~\citep{chen-rethinking-2017} is given as the reference, whereas the authentic source paper is ``Encoder-Decoder with Atrous Separable Convolution for Semantic Image Segmentation''~\citep{chen-encoder-2018}. The erroneously cited paper is actually the source paper of DeepLabv3, the previous version of DeepLabv3+. Another similar case in point is the published entity VQAv2. In one paper, its reference is given as ``VQA: Visual Question Answering''~\citep{antol-vqa-2015}, but in fact, this entity is proposed in ``Making the V in VQA Matter: Elevating the Role of Image Understanding in Visual Question Answering''~\citep{goyal-making-2017}. In addition to wrong references of completely different papers, some references are correct but the titles of the referred papers are mistakenly written. For instance, the source paper of the published entity EPIC-KITCHENS, ``Scaling Egocentric Vision: The EPIC-KITCHENS Dataset''~\citep{damen-scaling-2018}, is wrongly given as ``Scaling egocentric vision: the dataset''. Against such inaccurate citations, we make a strong call for enhanced awareness and stricter prevention to maintain the academic and scientific rigor.

Apart from inaccurate citations, we also find that many published entities are without citations. A large portion of them are established entities commonly accepted by computer science researchers, or at least within specific subfield communities. To further understand the situation of missing citations, we conduct a statistical analysis to figure out what types of these published entities are and how long it has been since they are proposed in their source papers.

\begin{figure}[ht]
\centerline{\includegraphics[width=0.9\textwidth]{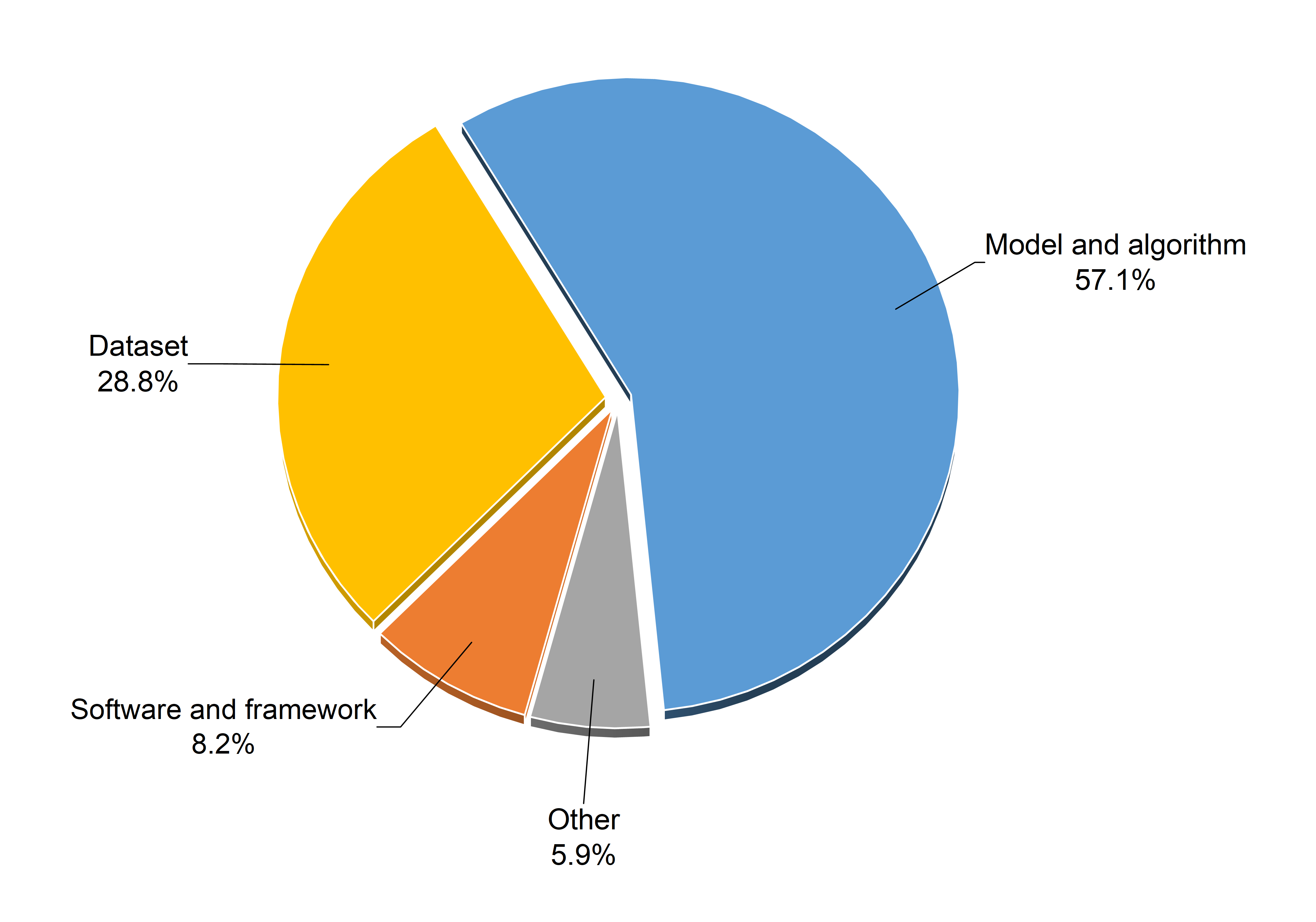}}
\caption{Distribution of types of published scientific entities with missing citations}
\label{fig:entity-type-distribution}
\end{figure}

First, we manually classify these published entities with missing citations into different types. The results are presented in Fig.~\ref{fig:entity-type-distribution}.

As shown in Fig.~\ref{fig:entity-type-distribution}, most of the entities with missing citations are models and algorithms, accounting for more than half of the total. Models and algorithms are at the core of computer science. The great proportion of models and algorithms with missing citations naturally speaks to their extensiveness in this field, i.e.\ most research in computer science is proposing models and algorithms as results. K-means is a fine example. This famous algorithm is presented by \citet{macqueen-some-1967} in 1967. Later, k-means become one of the basic knowledge that every researcher in the field of machine learning knows it. They use this algorithm to process data or develop new methods. But as it becomes overly popular, some researchers use it without citation. Datasets are also in the same situation. Over one-fourth of the published entities without citations are found to be datasets. Datasets are applied to test the performance of models and algorithms. Some datasets are used extensively and gradually become gold standards for certain tasks. In this process, some researchers take these gold standard datasets for granted and use them without citations. Two prominent examples are the Penn Treebank~\citep{marcus-building-1993} and LibriSpeech~\citep{panayotov-librispeech-2015}.

\begin{figure}[ht]
\centerline{\includegraphics[width=0.9\textwidth]{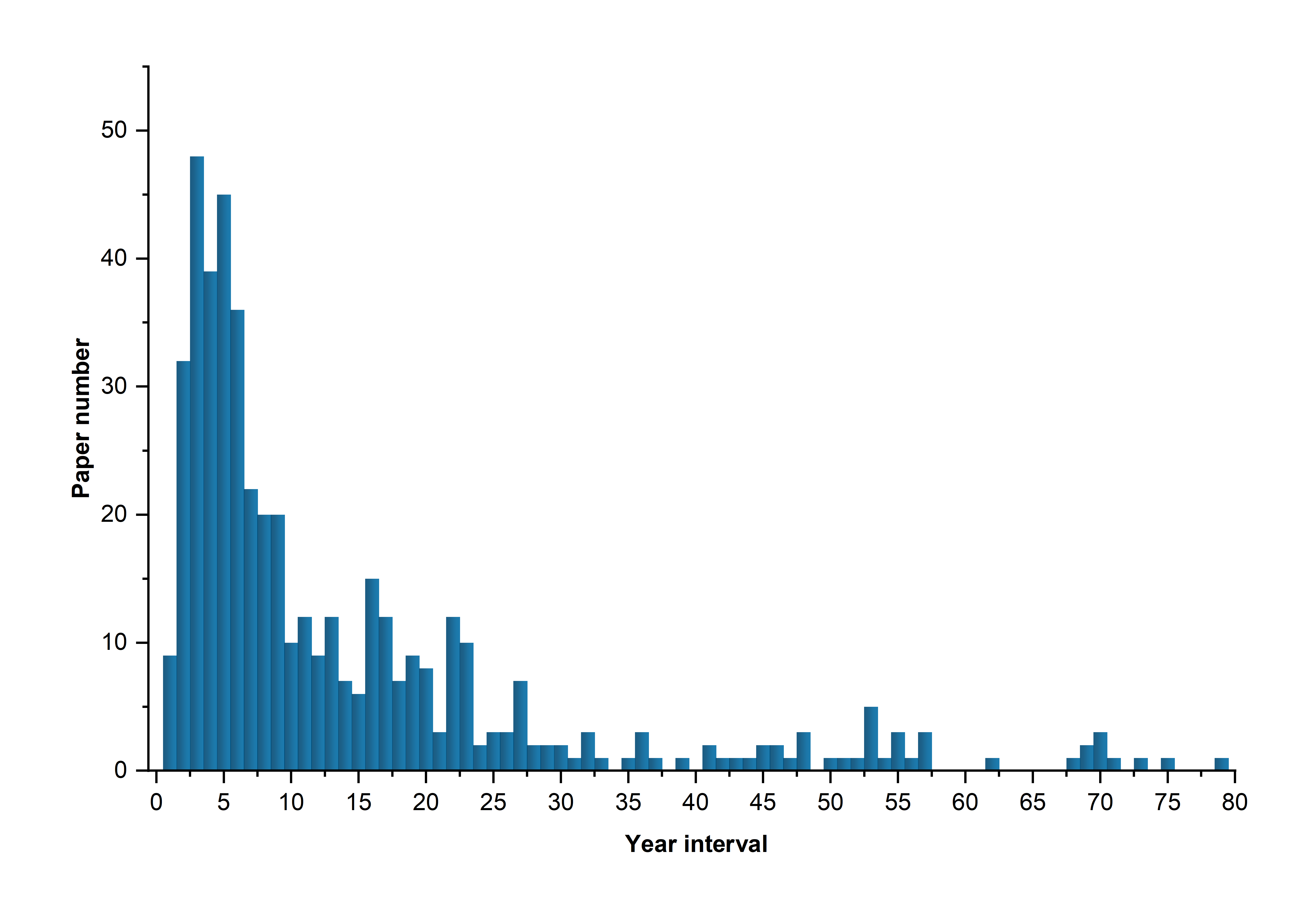}}
\caption{Histogram of the number of years source papers with missing citations have been published}
\label{fig:histogram-year}
\end{figure}

Then, we study how time makes a difference in the practice of missing citations for published entities. The source papers of the entities with missing citations are sorted in the light of how many years have passed since they are published. The results are presented in Table~\ref{tab:statistical-feature-year} with a histogram shown in Fig.~\ref{fig:histogram-year}.

\begin{table}[ht]
	\caption{Statistical features of the number of years source papers with missing citations have been published}
	\label{tab:statistical-feature-year}
	\centering
	\begin{tabular}{llllll}
		\hline\noalign{\smallskip}
		\textbf{Maximum}  &  \textbf{Minimum}  &  \textbf{Mode}  &  \textbf{Mean}  &  \textbf{Median}  \\
		\hline\noalign{\smallskip}
		79  &  1  &  3  &  13.9  &  8  \\
		\hline\noalign{\smallskip}
	\end{tabular}
\end{table}

Among the papers collected, the earliest published one is ``On Colouring the Nodes of a Network''~\citep{brooks-on-1941}, in which the published entity Brooks' theorem is proposed. Contrary to how we estimated before, some newly-proposed entities like SpecAugment also suffer from missing citations. The source paper of SpecAugment, ``SpecAugment: A Simple Data Augmentation Method for Automatic Speech Recognition''~\citep{park-specaugment-2019}, was published in 2019, just a year before the baseline year 2020 in our study. As can be seen in Fig.~\ref{fig:histogram-year}, a skewed distribution is registered with the distribution's peak off the center to the left. It means that not a few recent published entities suffer from missing citations. One explanation for this is that recent published entities are used more by researchers, whereas those entities published way before gradually lose their relevance. For the resistance of outliers, we take the median as a measure. The missing citations are among papers that have been published for 8 years. This time frame can be seen as the time needed for research results in computer science to grow into an existence of accepted knowledge. In these 8 years, these entities are first proposed, tested, then promoted while keeping on inspiring later researchers. They are used again and again, or they might even become textbook examples, to the point that some researchers are way too familiar with these entities to use them with citations.

\section{Conclusion}
\label{sec:conclusion}

In this paper, we propose a new method Citation Recommendation for Published Scientific Entity (CRPSE). Experimental results demonstrate that this method is effective in recommending source papers for published scientific entities. We further conduct a statistical analysis on missing citations of published scientific entities using CRPSE in the detection of missing citations. With analysis on open access papers from prestigious computer science conferences in 2020, we find that missing citations of published scientific entities are quite common, even for papers published in top computer science conferences. Most of the published scientific entities with missing citations are models and algorithms. As measured by the median, papers that propose published scientific entities with missing citations have been published for 8 years. This finding reveals the amount of time needed for a published scientific entity to develop into widely accepted knowledge to the point that researchers find it needless to give it citation in the domain of computer science. In view of the large amount of entities suffering from missing citations detected in our research, we call for proper citation practices for all published scientific entities to satisfy academic criteria.

\section*{Acknowledgments}

We would like to acknowledge the support of Yingmin Wang for improving the mathematical expressions. We are grateful to Li Lei, Xun Zhou, Lei Lin and Meizhen Zheng for their help in the data processing. We also appreciate two anonymous reviewers for their valuable comments. Special and heartfelt gratitude goes to the first author's wife Fenmei Zhou, for her understanding and love. Her unwavering support and continuous encouragement enable this research to be possible.

\section*{Funding}

This work is partly funded by the 13th Five-Year Plan project Artificial Intelligence and Language of State Language Commission of China (Grant No. WT135-38).

\section*{Compliance with ethical standards}

\textbf{Conflict of interest} The authors declare that there is no conflict of interest regarding the publication of this paper.


\bibliography{sn-article}


\end{document}